
\magnification =1200
\input harvmac
\overfullrule=0pt
\parskip=0pt plus 1pt

\def\np{Nucl. Phys.}
\def\pl{Phys. Lett.}
\def\prl{ Phys. Rev. Lett.}
\def\cmp{ Comm. Math. Phys.}

\def\gmunu{g_{\mu\nu}}
\def\ghat{{\hat g}^{\mu\nu}}
\def\gbar{{\hat{\bar g}}^{\mu\nu}}

\def\ino{{\psi}_{\mu\nu}}
\def\inoh{{\hat \psi}^{\mu\nu}}
\def\lc{{\cal L}_c}
\def\lgamma{{\cal L}_{\gamma}}
\def\lv{{\cal L}_v}
\def\ldv{{\cal L}_{d_P v}}
\def\lhat{{\hat{\cal L}}_v}
\def\Zalfa{Z_{\{n_{\alpha}\}}}
\def\mg{{\cal M}_g}

\def\ab{b_{\mu \nu}}
\def\abeta{\beta_{\mu \nu}}
\def\ci{{\cal C}^i}

{\nopagenumbers
\font\bigrm=cmb10 scaled\magstep1
\rightline{CERN-TH-7302/94}

\rightline{GEF-Th-6/1994}
\medskip
\centerline{\bigrm A FUNCTIONAL AND LAGRANGIAN FORMULATION}

\centerline{\bigrm {OF TWO DIMENSIONAL TOPOLOGICAL GRAVITY}\footnote{*}{This
work is partially supported by the ECRP, contract SC1-CT92-0789.}}

\vskip 1truecm
\centerline{C. M. Becchi, R. Collina}
\vskip 5pt
\centerline{\it Dipartimento di Fisica, Universit\`a di Genova}
\centerline{\it INFN, Sezione di Genova}
\centerline{\it Via Dodecaneso 33, I-16146 Genova, Italy}
\vskip 8pt
\centerline{C. Imbimbo\footnote{**}{On leave from INFN, Sezione
di Genova, Italy.}}
\vskip 5pt
\centerline{\it CERN, CH-1211, Geneva 23, Switzerland}
\bigskip
\centerline{ABSTRACT}

We reconsider two-dimensional topological gravity in a functional and
lagrangian framework. We derive its Slavnov-Taylor identities and
discuss its (in)dependence on the background gauge. Correlators of
reparamerization invariant observables are shown to be globally defined
forms on moduli space. The potential obstruction to their
gauge-independence is the non-triviality of the line bundle on moduli
space ${\cal L}_x$, whose first Chern-class is associated to the
topological invariants of Mumford, Morita and Miller.
\ \vfill

\leftline{CERN-TH 7302/94}

\leftline{GEF-Th 6/1994}

\leftline{June 1994}
\eject     }

\pageno=1

\beginsection 1. Introduction

Two-dimensional topological gravity \ref\lpw{J. Labastida, M. Pernici and
E. Witten, \np\ {\bf B310} (1988) 611.}\ is locally trivial as any topological
field theory and its "physical" content reduces to the coordinates of
the moduli space of the ``world-sheet'' Riemann surfaces. It is natural to
consider  these variables of infrared nature.
However, it has been shown \ref\vv{E. Verlinde and H. Verlinde, \np\ {\bf B348}
(1991) 457.}\ that the short distances play a crucial and unexpected
r\^ole in topological gravity. It turns out that, for a suitable
parametrization of the moduli space, almost all the correlation functions of
the known "observables" of the theory can be made to  vanish at the
interior points of this space. Their only non-trivial contributions come
from contact terms with the possible nodes of the ``world-sheet'', that is from
the boundary of moduli space. This looks very much like an ultraviolet
phenomenon.

To have a better understanding of this double  nature of the correlation
functions it is necessary  to use a local formulation of the theory.
This motivates the present paper, which seeks to define a field
functional framework for two-dimensional topological gravity.

With the same purpose in mind, we have discussed
and identified in \ref\bci{C.M. Becchi, R. Collina and C. Imbimbo,
\pl {\bf B 322} (1994) 79.}\ the field
structure of a class of "observables" that are  covariant under
reparametrizations of the ``world-sheet'' and that correspond to the
known relevant operators of the theory. To compute their correlation
functions it is necessary to identify a non-degenerate action for the
theory. The existence of $6g-6$ moduli parametrizing the possible
inequivalent complex structures of surfaces of genus $g$ automatically
induces the existence of the same number of zero-modes for each
antighost field. It is therefore necessary to introduce a suitable gauge
fixing for these zero-modes.

After the introduction of the suitable gauge fixing,  the theory is
characterized by a family of Slavnov-Taylor identities exhibiting its
BRS invariance. We show that, in the case of BRS invariant operators
that are independent of the antighost zero-modes, these identities
signify that the correlation functions are closed forms on the moduli
space. To understand if these correlation functions are globally defined
we study their properties under arbitrary variations of the gauge slice.
We show that in general such variations change the correlation functions
by terms which are locally exact forms on the moduli space. However these
terms vanish in the case of modular transformations constant on moduli
space. From this we conclude that the correlation functions are globally
defined forms on
moduli space when one chooses gauge slices whose transition functions
are modular transformations. We also point out that a potential
obstruction to the background gauge independence of such globally
defined forms is the non-triviality of the line bundle ${\cal L}_x$
associated to the topological invariants of Mumford, Morita and Miller
\ref\mumford{D. Mumford, ``Towards and enumerative geometry of the
moduli space of curves'', Arithmetic and Geometry, Michael Artin and
John Tate, eds., Birkh\"auser, Basel, 1983.}-\nref\morita{S. Morita,
Invent. Math. {\bf 90} (1987) 551.}\nref\miller{E. Miller, J. Diff. Geo. {\bf
24}(1986) 1.}\ref\wit{E. Witten, Supplement to the Journal of
Differential Geometry, {\bf 1} (1991) 243.}.

The paper is thus organized: in the following section 2, we revisit the
known observables of the theory; Section 3 contains the construction of the
Lagrangian;  the Slavnov Taylor identities are discussed in Section 4; and
the variations of the gauge-slice in section 5.

\beginsection 2. The Observables

Two-dimensional topological gravity \lpw\ is a
topological quantum field theory characterized by the following BRS
transformation laws:
\eqn\brs{\eqalign{s \gmunu &= \lc \gmunu + \ino \cr
s \ino &= \lc \ino - \lgamma \gmunu ,\cr}}
where $\gmunu$ is the two-dimensional metric, $\ino$is the gravitino
field, $c^{\mu}$ is the ghost vector field and $\gamma^{\mu}$ is
the superghost vector field. $\lc$ and $\lgamma$ denote the action of
infinitesimal diffeomorphisms with parameters $c^{\mu}$ and
$\gamma^{\mu}$ respectively.

A class of observables local in the fields $\gmunu ,\ino ,c^{\mu}$
and $\gamma^{\mu}$ can be constructed \vv ,\bci\ starting from the Euler
two-form
\eqn\euler{\sigma^{(2)} = {1\over 2}\sqrt{g} R \epsilon_{\mu \nu}
dx^{\mu}\wedge dx^{\nu},}
where $R$ is the two-dimensional scalar curvature and $\epsilon_{\mu \nu}$
is the antisymmmetric numeric tensor defined by $\epsilon_{12}=1$.
Since $s$ and the exterior differential $d$ on the two-dimensional
world-sheet commute among themselves, the two-form in Eq.\euler\ gives rise
to the descent equations:
\eqn\descent{\eqalign{
s \sigma^{(2)} =& d \sigma^{(1)}\cr
s \sigma^{(1)} =& d \sigma^{(0)} \cr
s \sigma^{(0)} =& 0. \cr}}
The zero-form $\sigma^{(0)}$ and the one-form $\sigma^{(1)}$
are computed to be
\eqn\observables{\eqalign{
\sigma^{(0)} =& \sqrt{g} \epsilon_{\mu \nu} \left[{1\over 2}c^{\mu}
c^{\nu}R + c^{\mu}D_{\rho}(\psi^{\nu\rho}
- g^{\nu\rho}\psi^{\sigma}_{\sigma}) + D^{\mu}\gamma^{\nu} -{1\over
4}\psi^{\mu}_{\rho}\psi^{\nu\rho}\right] \cr
\sigma^{(1)} =& \sqrt{g}\epsilon_{\mu\nu}\left[c^{\nu} R +
D_{\rho}(\psi^{\nu \rho} - g^{\nu\rho}\psi^{\sigma}_{\sigma})
\right]dx^{\mu}.\cr}}

The Euler form, being a topological invariant, is locally $d$-exact,
\eqn\exact{\sigma^{(2)}= d \omega^{(1)}.}
This, together with the descent equations \descent , implies that
\eqn\exactbis{\eqalign{
\sigma^{(1)} =&d\omega^{(0)} + s\omega^{(1)}\cr
\sigma^{(0)} =&s\omega^{(0)}.\cr}}
Thus $\sigma^{(0)}$ is locally BRS trivial. However, since $\omega^{(1)}$
cannot be chosen to be a globally defined
1-form, it follows from Eqs. \exactbis\ that $\omega^{(0)}$ cannot be chosen
to be a globally defined scalar field either. This means that $\sigma^{(0)}$
is a non-trivial class in the cohomology of $s$ acting on the space of the
{\it reparametrization covariant} tensor fields. One can verify explicitly
that such cohomology is in one-to-one correspondence with the
{\it semi-relative} state BRS cohomology defined on the state space
of the theory quantized on the infinite cylinder in the conformal
gauge. In fact, choosing a complex structure $\mu$ on the two-dimensional
surface, with $(ds)^2\sim |dz +\mu d{\bar z}|^2$, one obtains
\eqn\holo{\eqalign{
\omega^{(1)} =& {2\over \Theta}\left[\partial \mu - \mu{\bar
\partial}{\bar \mu}+ {1\over 2}({\bar \nabla}-\mu
\nabla )\ln{\Theta}\right]d{\bar z} - c.c.\cr
\omega^{(0)} =& \nabla c^z +{\bar c}^{\bar z}\omega_{{\bar z}}^{(1)} + {{\bar
\mu}\psi_z \over \Theta} - c.c., \cr}}
where $\Theta \equiv 1 -\mu{\bar \mu}$, $\nabla \equiv \partial
-\mu{\bar \partial}$, $c^z$ and $\psi_z$ are the holomorhic components of
the ghost field and of the traceless part of the gravitino fields,
and $c.c.$ denotes the complex conjugate expression
in which all quantities are substituted with their barred expressions.
In the (super)-conformal gauge,
\eqn\conformal{\omega^{(0)} = \partial c -{\bar \partial}{\bar c},}
and, at the level of states:
\eqn\states{\omega^{(0)}(0)|0> = c_{0}^{-}|0>,}
where $|0>$ is the $SL(2,C)$ invariant conformal vacuum.
Thus the state created by the operator $\sigma^{(0)}(0)$ --- the so-called
``dilaton'' state --- is non-trivial in the cohomology of
the BRS operator acting on the space of states annihilated by $b_0^{-}$:
\eqn\dilaton{\sigma^{(0)}|0> = s (c_0^{-}|0>).}
This is known, in the operator formalism, as the semi-relative BRS state
cohomology \ref\nelson{P. Nelson, \prl\ {\bf 62} (1989) 993.},
\ref\distler{J. Distler and P. Nelson, \cmp\ {\bf 138} (1991) 273.}.

Since the superghosts $\gamma^{\mu}$ are commutative, one can build an
infinite tower of cohomologically non-trivial operators by taking
arbitrary powers of $\sigma^{(0)}$:
\eqn\osser{\sigma_n^{(0)} \equiv (\sigma^{(0)})^n}
with $n=0,1,\ldots$ The corresponding 2-forms
\eqn\obtwon{\sigma_n^{(2)} = n (\sigma_n^{(0)})^{n-1} \sigma^{(2)}
+ {n(n-1)\over 2}(\sigma_n^{(0)})^{n-2}\sigma_n^{(1)}\wedge
\sigma_n^{(1)}}
all belongs in the $s$-cohomology modulo $d$ on the space of the
reparametrization covariant tensor fields.

\beginsection 3. The Lagrangian

In order to evaluate correlators of observables $\sigma_n$, the
choice of a lagrangian is required. The theory being topological, the
choice of a lagrangian amounts to fixing the gauge.

Let $\mg$ be the moduli space of two-dimensional Riemann surfaces
of a given genus $g$, and let $m=(m^i)$, with $i=1,\ldots, 6g-6$,
be local coordinates on $\mg$. Fixing the gauge means choosing a
background metric ${\bar g}_{\mu\nu}(x;m)$ for each gauge
equivalence class of metrics corresponding to the point $m$ of $\mg$.

It is convenient to decompose ${\bar g}_{\mu\nu}$ as follows:
\eqn\background{{\bar g}_{\mu\nu}(x;m) \equiv \sqrt{g}
{\hat{\bar g}}_{\mu\nu}(x;m) \equiv e^{{\bar\varphi}}
{\hat{\bar g}}_{\mu\nu}(x;m),}
with $det({\hat{\bar g}})_{\mu\nu}=1$. ${\hat g}_{\mu\nu}$ is given by the
analogous definition for $\gmunu$. We also introduce the
the tensor density,
$$\inoh \equiv \sqrt{g}(\psi^{\mu\nu} -\half g^{\mu\nu}\psi^{\sigma}_{\sigma}),
$$
\noindent in correspondence with the traceless part of the gravitino field.
${\bar\gmunu}$ defines a gauge-slice on the field space whose associated
lagrangian reads as follows:
\eqn\lagra{
{\cal L} = s \left[\half b_{\mu \nu} (\ghat -\gbar) +
\half \beta_{\mu \nu}(\inoh - d_P\gbar ) + \chi \partial_{\mu}(\ghat
\partial_{\nu}(\varphi -{\bar \varphi}))\right].}
In Eq.\lagra\ we have introduced the ``exterior derivative''
operator
$$d_P \equiv p^i {\partial \over \partial m^i},$$
\noindent where $p^i$ are the anti-commuting supermoduli, with $i=1,\ldots
6g-6$, the superpartners of the commuting moduli $m^i$.
$\ab$, $\abeta$ and $\chi$
are the anti-ghost fields, with ghost numbers $-1,-2$ and $0$
respectively. Their BRS transformation laws are given by
\eqn\multipliers{\eqalign{
s\,\ab = &\Lambda_{\mu \nu}, \qquad s\,\Lambda_{\mu \nu} = 0 \cr
s\,\abeta = &L_{\mu \nu}, \qquad s\,L_{\mu \nu} = 0 \cr
s\,\chi = &\lc \chi + \pi, \; s\,\pi = \lc \pi -\lgamma \chi,\cr }}
where $\Lambda_{\mu\nu}$, $L_{\mu\nu}$ and $\pi$ are Lagrangian
multipliers.
The Lagrangian in Eq.\lagra\ written out in extended form reads:

\eqn\lagratwo{\eqalign{
{\cal L} =& \half \Lambda_{\mu \nu}(\ghat -\gbar) + \half L_{\mu \nu}
(\inoh  - d_P\gbar ) \cr
& - \half \ab \lc \ghat  - \half \abeta \lgamma \ghat \cr
& +\half \inoh \left[ (\lc\beta)_{\mu \nu} + \ab + 2\partial_{\mu}\chi
\partial_{\nu}(\varphi -{\bar \varphi})\right]\cr
&+\pi \partial_{\mu} (\ghat \partial_{\nu} (\varphi -{\bar
\varphi}))
- \chi \partial_{\mu} (\ghat \partial_{\nu}\psi^{\prime}) ,\cr}}
where
\eqn\psprime{\psi^{\prime} \equiv {\bar D}_{\sigma} c^{\sigma} +
\half\psi^{\sigma}_{\sigma}.}

Eq.\lagratwo\  shows the standard form of a topological Lagrangian that,
however,
in the present case is degenerate due to the presence of $6g-6$ moduli.
Indeed,
$\beta^{(i)}\equiv \int \abeta {\partial \over \partial m^i}\gbar$,
define $6g-6$ zero-modes of the antighost field  $\beta$.

To remove this degeneracy we must introduce further gauge fixing
terms. The natural way of generating these terms is to extend the definition
of the BRS generator $s$, assuming the following BRS transformation laws
for the moduli and super-moduli:
\eqn\moduli{\eqalign{
s\, m^i = &\ci ,\qquad s\,\ci = 0 \cr
s\, p^i = &-\Gamma^i , \;\; s\,\Gamma^i =0, \cr}}
where $\ci$ and $\Gamma^i$ are respectively anti-commuting and
commuting Lagrange multipliers.

After this extension of the BRS transformations four new terms
appear in the lagrangian:
\eqn\lagrathr{ \half\abeta d_{\Gamma}\gbar + \half\ab d_C \gbar +
\half\abeta d_P d_C \gbar + \chi \partial_{\mu}(\ghat
\partial_{\nu}d_C{\bar\varphi}), }
where the notation $d_C\equiv \ci {\partial \over \partial m^i}$ and
$d_{\Gamma}\equiv \Gamma^i {\partial \over \partial m^i}$ has been
introduced.
The first term in Eq.\lagrathr\  is the wanted zero mode
fixing term. The second term generates the compensating term for it
determinant.

After this second gauge fixing, integrating out the Lagrangian multipliers
$\Lambda_{\mu \nu}$, $L_{\mu\nu}$, $\pi^{\prime}$ and $\chi$
forces the metric and the gravitino field to take their background
values,
\eqn\substitutions{\eqalign{
\ghat &\rightarrow \gbar ,\qquad \varphi\rightarrow{\bar\varphi},\qquad
\inoh \rightarrow d_P \gbar \cr
\half\psi^{\sigma}_{\sigma}& + {\bar D}_{\sigma}c^{\sigma}\rightarrow
d_C{\bar\varphi},\cr}}
and the lagrangian becomes
\eqn\newlagra{\eqalign{
{\cal L}^{\prime} =&\half ( -\ab \lc \gbar - \abeta \lgamma \gbar
+d_P\gbar (\lc \beta)_{\mu\nu}\cr
&+\ab (d_C \gbar - d_P \gbar ) + \abeta d_{\Gamma}\gbar +
\abeta d_P d_C \gbar ).\cr}}

In the following we will repeatedly make use of the fact that, when
the observables do not contain the anti-ghost zero modes $b^{(i)}$,
integrating them out introduces into the correlators the factor
\eqn\simplela{\prod_{i=1}^{6g-6} \delta (\ci - p^i).}

If moreover there are no antighost zero modes $\beta^{(i)}$ and no
antighost fields $\ab$ in the observables, one can integrate out
$\beta^{(i)}$ as well. This produces a further factor
\eqn\betaout{\prod_{i=1}^{6g-6} \delta (\Gamma^i).}

\beginsection 4. B.R.S. Identities

Let us now consider expectation values of observables $O_{n_{\alpha}}
(\Phi)\equiv \int \sigma_{n_{\alpha}}^{(2)}$, where by $\Phi$ we denote
collectively all the quantum fields but the moduli and the super-moduli
$m^i$ and $p^i$. It is useful to consider functional averages in which
one integrates only with respect to the quantum fields $\Phi$ :
\eqn\averages{\eqalign{
Z_{\{ n_{\alpha}\}}(m^i,p^i) \equiv &\int\,[d\Phi] e^{-
S(\Phi ;m^i,\,p^i)}\prod_{\alpha} O_{n_{\alpha}}(\Phi)\cr
\equiv &<\prod_{\alpha} O_{n_{\alpha}}(\Phi )>. \cr}}

Because of ghost number conservation, $\Zalfa$ is a monomial of the
anti-commuting super-moduli:
\eqn\ghost{\Zalfa (m^i,p^i)  = Z_{i_1\ldots i_N}(m^i)
p^{i_1}\ldots p^{i_N},}
where $N$ is the total ghost number of the observables $O_{n_{\alpha}}$:
\eqn\ghnumber{N = \sum_{\alpha} ({\rm ghost\#}\, O_{n_{\alpha}})
= 2 \sum_{\alpha} (n_{\alpha} -1).}
Under a reparametrization ${\tilde m}^i = {\tilde m}^i(m)$ of the
local coordinates $m^i$ on the moduli space $\mg$, the supermoduli transform
as follows:
$${\tilde p}^i = {\partial {\tilde m}^i \over \partial m^j}p^j.$$
One can therefore identity the anti-commuting supermoduli with the
differentials on the moduli space, i.e. $p^i \rightarrow dm^i.$
Correspondingly, the function $\Zalfa (m^i,p^i)$ of the moduli and
super-moduli can be thought of as a N-form on the moduli space $\mg$, at least
{\it locally} on $\mg$. The question of whether or not such local form
extends to a globally defined form on $\mg$, will be addressed shortly.

Assume for the moment that form  $\Zalfa (m)$ is globally defined on
$\mg$. Whenever the following ghost number selection rule is satisfied,
\eqn\selection{N = 2 \sum_{\alpha} (n_{\alpha} -1) = 6g-6,}
$\Zalfa (m)$ defines a measure on $\mg$ which can be integrated
to produce some number. The collection of these  numbers encode the
gauge-invariant content of two-dimensional topological gravity.

It is easy to show that the action of BRS operator $s$ on the quantum
fields $\Phi$ translates into the action of the exterior differential $d_P
\equiv p^i\partial_i$ on the forms $\Zalfa (m)$. More precisely,
one can prove the following BRS identities:
\eqn\identities{\eqalign{
(i)\;\;  &s\, O_{n_{\alpha}}(\Phi) = 0 \Rightarrow d_P \Zalfa (m)= 0\cr
(ii)\;\; &O_{n_{\alpha}} = s\, X_{n_{\alpha}}(\Phi) \Rightarrow
\Zalfa (m) = d_P W(m)\cr}}
where $W(m) \equiv <X_{n_{\alpha}} \prod_{\beta \ne\alpha}
O_{n_{\beta}}>$, with the provision that neither $O_{n_{\alpha}}$
nor $X_{n_{\alpha}}$ contains the anti-ghost zero modes and the
antighost field $b$.

The proof of $(i)$, for example, goes as follows:
\eqn\closed{\eqalign{
d_P \Zalfa &= <- d_P S \prod_{\alpha} O_{n_{\alpha}}(\Phi)> \cr
&= <- d_C S \prod_{\alpha}O_{n_{\alpha}}(\Phi)> \cr
&= <\sum_{\Phi} s\,\Phi{\delta S \over \delta \Phi}\prod_{\alpha}
O_{n_{\alpha}}(\Phi)> \cr
&= <s \prod_{\alpha} O_{n_{\alpha}}(\Phi)> = 0,\cr}}
since, under the stated conditions, one can perform the
substitutions  $\ci \rightarrow p^i$ and $\Gamma^i \rightarrow 0$
(see Eqs.\simplela\ and \betaout).
The proof of $(ii)$ is analogous.

\beginsection 5. (In)dependence on the gauge and global properties

We would like to study the dependence of $\Zalfa (m)$ on the
gauge-fixing function ${\bar g}_{\mu\nu}(x;m)$ which specifies for each point
$m$ of the moduli space $\mg$ a representative two-dimensional metric.

Given the gauge-slice corresponding to ${\bar\gmunu}(x;m)$, a different gauge
choice corresponding to ${{\bar \gmunu}}^{\prime}(x;m)$ will be related to
${\bar \gmunu}$ by a diffeomorphism $\xi^{\mu}(x;m)$ depending on $m$:
\eqn\diff{
{\bar g}^{\prime}_{\mu\nu}(x;m) = {\partial \xi^{\rho}\over \partial
x^{\mu}}{\partial \xi^{\sigma}\over \partial x^{\nu}}{\bar g}_{\rho
\sigma}(x;m).}

Consider for simplicity the case of an infinitesimal diffeomorphism
$\xi^{\mu}\sim x^{\mu}+ v^{\mu}(x;m)$, for which
\eqn\infinitesimal{{\bar g}^{\prime}_{\mu\nu} (x;m) \sim
{\bar g}_{\mu\nu}(x;m) + (\lv{\bar g})_{\mu\nu}(x;m).}
Denoting by $W_v$ the action of the infinitesimal diffeomorphism in
Eq.\infinitesimal\ on the form $\Zalfa$, one has
\eqn\action{\eqalign{
&W_v \Zalfa = <\int s\;\half \left[\ab (\lv {\hat{\bar g}})^{\mu\nu} +
\abeta d_P (\lv {\hat{\bar g}})^{\mu\nu} + 2\chi \partial_{\mu}(\ghat
\partial_{\nu}\lv {\bar \varphi})\right] \prod_{\alpha}
O_{n_{\alpha}}>\cr
&= <s \int \half\left[\ab (\lv {\hat{\bar g}})^{\mu \nu} - (\lv
\beta)_{\mu\nu}d_P\gbar + \abeta (\ldv {\hat{\bar g}})^{\mu \nu}
+2\chi \partial_{\mu}(\ghat\partial_{\nu}\lv {\bar \varphi})\right]
\prod_{\alpha}O_{n_{\alpha}}>\cr
&= <s\int I_v S \prod_{\alpha}O_{n_{\alpha}}>, \cr}}
where $\ldv \equiv p^i {\cal L}_{\partial_i v}$ and where we introduced
the operator $I_v$ acting on the fields $\Phi$ as
follows:
\eqn\contraction{\eqalign{
I_v c^{\mu} &= v^{\mu}(x;m)\cr
I_v \gamma^{\mu} &= -d_P v^{\mu}(x;m)\cr
I_v \Phi &= 0 \qquad \hbox{for the other fields.} \cr}}
By considering the anticommutator of $I_v$ and $s$,
\eqn\anti{\lhat = \{ s , I_v \},}
one obtains an operator $\lhat$ acting on a generic field
supermultiplet $\Phi \equiv (\phi, {\hat \phi})$ as follows:
\eqn\diffhat{\eqalign{
\lhat \phi =& \lv \phi \cr
\lhat {\hat \phi} =& \lv {\hat \phi} +\ldv \phi .\cr}}

In terms of $I_v$ and $\lhat$, the Ward identity in Eq.\action\
becomes:
\eqn\ward{\eqalign{
W_v \Zalfa =& <s\; I_v \prod_{\alpha} O_{n_{\alpha}}>
=<\lhat \prod_{\alpha} O_{n_{\alpha}}> \cr
= & d_P < I_v \prod_{\alpha} O_{n_{\alpha}}>,\cr}}
where in the last line we used $(ii)$ of Eq.\identities .
Eq.\ward\ is the fundamental equation to understand both the global
properties of $\Zalfa$ and its gauge dependence.

Let us consider first the question of the global definition of $\Zalfa$.
What makes this property not obvious is that ${\bar\gmunu}(x;m)$ cannot be
chosen to be a continuous function of $m$. In fact ${\bar\gmunu}(x;m)$ is
a section of the bundle over $\mg$ defined by the space of
two-dimensional metrics of given genus. This bundle is not trivial,
and therefore
it does not admit a global section. On some coordinate patch on $\mg$,
${\bar\gmunu}$ must jump to a ${\bar\gmunu}^{\prime}$ related to
${\bar\gmunu}$ by
a diffeomorphism $\xi^{\mu}(x;m)$ as in Eq.\diff . From Eq.\ward\
it follows that, for a generic local section ${\bar\gmunu}$, $\Zalfa$ jumps as
well, and hence is not globally defined. However, the bundle in
question is non-trivial only because of modular transformations. This means
that it is possible to choose sections ${\bar\gmunu}$ whose transition
functions $\xi^{\mu}(x;m)$ are $m$-independent modular transformations.

Eq.\diffhat\ implies that for infinitesimal diffeomorphisms $v^{\mu}(x;m)$
which are independent on $m$, the operator $\lhat$ reduces to
usual diffeormophisms, i.e.
\eqn\constant{\partial_i v^{\mu}=0 \Rightarrow \lhat = \lv.}
Thus, under such diffeomorphisms and for observables which are
reparametrization invariant, $\Zalfa$ is invariant:
\eqn\invariant{W_v \Zalfa = <\lv \prod_{\alpha}O_{n_{\alpha}}>= 0.}
It is not difficult to see that this remains true for finite,
$m$-independent diffeormophisms $\xi^{\mu}$.

In conclusion, the Ward identity in Eq.\ward\ ensures that if we
consider sections ${\bar\gmunu}$ whose transition functions are $m$-independent
modular transformations and restrict ouselves to reparametrization invariant
observables,
the corresponding $\Zalfa$ are globally
defined forms on the moduli space $\mg$.  For the same reason,
observables like $\sigma^{(0)} = s\,\omega^{(0)}$ which are locally
BRS trivial, give rise to forms $<\sigma^{(0)}\ldots > =d_P
<\omega^{(0)}\ldots >$ which are locally
but not globally exact on $\mg$. This is the reason why
the relevant BRS cohomology is the BRS cohomology acting on the
space of reparametrization covariant tensor fields. As we have seen,
this cohomology precisely corresponds to the
semi-relative $b_0^{-}$ BRS state cohomology of the operator formalism.

Eq.\ward\ also implies that two different sections ${\bar\gmunu}$ and
${\bar\gmunu}^{\prime}$, whose transition functions are modular
transformations,
give rise to globally defined forms $\Zalfa$ and ${\Zalfa}^{\prime}$ which
differ by a locally exact form on $\mg$,
\eqn\esatto{\Zalfa - {\Zalfa}^{\prime} = d_P <I_v \prod_{\alpha}
O_{n_{\alpha}}>,}
where $v = (v^{\mu}(m;x))$ is the infinitesimal diffeomorphism relating
${\bar\gmunu}$ to ${\bar\gmunu}^{\prime}$. Gauge independence of the
theory would require that the integral over $\mg$ of the form in the
r.h.s. of Eq.
\esatto\ vanishes. Thus, the question of independence on the background
metric ${\bar g}_{\mu\nu}$ is the one of the global
definition of the form $<I_v \prod_{\alpha} O_{n_{\alpha}}>$.

The vector field $v^{\mu}(m;x)$, for a fixed point $x$ on the
world-sheet, is a section of the complex line bundle ${\cal L}_x$ on
$\mg$, whose fiber is the cotangent space of the world-sheet at $x$.
This bundle is non-trivial, and its first Chern-class $c_1({\cal L}_x)$
is related to the topological invariants of Mumford, Morita and Miller
\mumford -\wit . It follows that $c_1({\cal L}_x)$ is precisely the
obstruction to
choose a globally defined $v^{\mu}(m;x)$. Since $v^{\mu}(\mu;x)$ is not
globally defined, the form in the r.h.s. of Eq.\esatto\ is potentially
not globally exact.

In this sense, the non-trivial content of two-dimensional topological
gravity is captured by the ``anomaly'' Equation \ward . The work of
Verlinde and Verlinde \vv\ suggests that there exists a suitable
redefinition of the functional measure of two-dimensional gravity
at the boundary of the moduli space $\mg$ which restores background gauge
independence while preserving BRS invariance. We hope to come back to
this issue in the future.

\beginsection Acknowledgements

It is a pleasure to thank R. Stora for interesting discussions.

This work is carried out in the framework of the European Community Research
Programme ``Gauge theories, applied supersymmetry and quantum gravity'', with a
financial contribution under contract SC1-CT92-0789.

\listrefs
\bye